\documentclass[preprint,12pt,authoryear]{article}

\usepackage{geometry}
\usepackage{array}
\usepackage{graphicx} 
\usepackage{longtable} 
\usepackage{amsmath}
\usepackage{amssymb}
\usepackage{comment} 
\usepackage{xcolor}
\usepackage{fancyvrb}
\usepackage{doi}
\usepackage{cleveref}
\usepackage{acro}
\usepackage{siunitx}
\usepackage[style=phys, backend=biber, maxbibnames=4,maxcitenames=2,]{biblatex}
\usepackage{placeins}

\usepackage[utf8]{inputenc}
\usepackage{hyperref}
\hypersetup{
	unicode,
	pdfauthor={Author One, Author Two, Author Three},
	pdftitle={\texttt{MuDirac} 1.3.0: A Sustainable Software Tool for Calculating Ground State Nuclear Properties Using Muonic X-Ray Measurements},
	pdfsubject={A simple article template},
	pdfkeywords={negative muons, muonic x-rays, nuclear properties},
	pdfproducer={LaTeX},
	pdfcreator={pdflatex}
 }

\title{\texttt{MuDirac} 1.3.0: A Sustainable Software Tool for Calculating Ground State Nuclear Properties Using Muonic X-Ray Measurements}
\author{L. Liborio$^1$\footnote{Corresponding author: leandro.liborio@stfc.ac.uk} \and   M. Kumar$^1$ \and  S. Devadasan$^1$ \and P. D. Jones$^2$ \and   M. Plummer$^1$ \and A. Hillier$^4$ \and   A. P. Bartok$^2$ $^3$}

\date{
	$^1$ Scientific Computing Department, STFC, UKRI, Harwell, UK. \\%
	$^2$Department of Physics, University of Warwick, Coventry, UK. \\%
    $^3$Warwick Centre for Predictive Modelling, School of Engineering, University of
Warwick, Coventry, UK. \\
    $^4$ISIS Facility, STFC, UKRI, Harwell, UK.
}
\newcommand{\dd}{\textrm{d}}
\begin{document}

\maketitle
	
	\begin{abstract}
    The nuclear charge radius is one of the most fundamental quantities of the atomic
    nucleus. It can be
    deduced from a combination of experimental measurements of muonic X-ray transition energies with modelling of those X-ray transition energies.  In this work we present  \texttt{MuDirac} (1.3.0), which is an open,
    publicly available, sustainable and computationally efficient software tool that will be at put the
    disposal of the negative muon community.  With  \texttt{MuDirac} (1.3.0), the community will be able to accurately and efficiently estimate
    nuclear properties, such as the nuclear charge radius, by assuming a 2-parameter Fermi distribution of the nuclear charge.
	\end{abstract}

\section{Introduction}

Negative muons have a mass of around 200 times the mass of an electron. Hence, they can be thought of as ``heavy electrons'' that, when implanted in a sample, replace an electron in the outer shell of an atom and then decay through its corresponding energy states.  Each transition between these energy states produces an X-ray characteristic of the atom that captured the muon.  While there exist tabulated databases of muonic atom X-ray transition energies\cite{Zinatulina2018}, the Muon Spectroscopy Computational Project (MSCP)\footnote{https://muon-spectroscopy-computational-project.github.io/} -a software development project based at Science and Technology Facilities Council in the UK- is developing \texttt{MuDirac}\cite{sturniolo2021}, a modern, open-source, sustainable and publicly available software tool that is capable of computing muonic X-ray spectra by solving the radial Dirac equation using a radial potential $V(r)$. \texttt{MuDirac} can compute muonic atom transition energies up to precisions on the order of the keV, accounting for the effects of finite nuclear size, vacuum polarizability, non-relativistic recoil corrections and electronic screening.  \newline

The nuclear charge radii is a
fundamental property of the atomic nucleus, and its value depends on features of the nuclear structure, such as a nuclear deformation and nuclear charge distribution.  What is considered the nuclear charge radii is the root mean square (rms) radius, which is defined using the nuclear charge distribution $\rho$ as:

\begin{equation}
\label{rms}
    R_{\textrm{rms}}=\sqrt{{\overline{r^{2}}}}= \left[ \frac{ \int r^{2}\rho(r)\dd r }{\int \rho(r)\dd r} \right]^\frac{1}{2}
\end{equation}

The nuclear charge radius is a key property that has been systematically measured for almost all stable nuclei\cite{marinova_2013} and, recently, also for some unstable nuclei\cite{Suda2017}. This key property can be deduced from the X-ray transition energies of a muonic atom\cite{FRICKE1995}. The binding energy of the muonic atom is sensitive to the charge distribution of the nucleus and, in principle, the nuclear charge radius can be determined  by using the muonic X-ray measurements. There is a complex relationship between the muonic X-ray transition energies and the nuclear charge distribution $\rho$. A possible way of treating this problem is assuming a functional form for $\rho$, such as the 2-parameter Fermi distribution (2pF)\cite{Hofstader1958}: 

\begin{equation}
    \label{2pf}
    \rho(r)=\rho_{0} \left[ 1 + \exp\left({4\ln(3)\frac{(r-c)}{t}} \right)  \right]^{-1}
\end{equation}

\begin{figure}[!htb]
    \centering
    \includegraphics[width=1\linewidth]{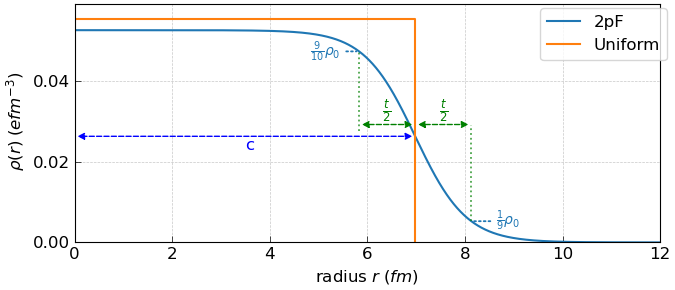}
    \caption{Charge density in the spherically symmetric Uniform model and 2pF model for $^{197}\text{Au}$. The uniformly charged nucleus as  has a radius described by $R_u = 1.2 \times A^{\frac{1}{3}}$, where $A$ is the nucleon number for $^{197}\text{Au}$. The 2pF model here uses $t=\qty{2.3}{fm}$ and $c$ is annotated to show that it represents the half density radius in the 2pF model. This happens to coincide with $R_u$ in the Uniform model but this is generally not the case. The 2pF model does not have a hard edge like the uniform model. Instead, the edge is diffuse and $\rho$ falls from  $\frac{9}{10}\rho_0$ to $\frac{1}{9}\rho_0$  across distance $t$ around $r=c$ as shown by the annotations. }
    \label{fig: nuclear models rho 1D}
\end{figure}

Equation (\ref{2pf}) describes a spherically symmetric nuclear charge distribution that comprises a uniformly charged spherical nucleus with a `skin' of thickness $t$, which is defined as the distance at which the charge density reduces from 90$\%$ to 10$\%$ of the density at the centre of the uniformly charged nucleus.  The other parameter of the 2pF model is $c$, which describes the half-density radius: the distance from the centre of the spherical nucleus where the charge density is half of that of the uniform charge at the centre. Figure (\ref{fig: nuclear models rho 1D}) compares the nuclear charge distributions in the uniform and 2pF models for an example atom $^{197}\text{Au}$. Using the 2pF model allows for a soft edge to the nuclear charge density. 

In the current version of \texttt{MuDirac} (1.2.4), 
the 2pF model uses $t=\qty{2.3}{fm}$, while $c$ is obtained from nuclear data tables\cite{marinova_2013}. Hence, the $c$ and $t$ parameters of the 2pF model are defined \textit{a priori}.  This is because the initial focus of \texttt{MuDirac} was the calculation of muonic X-ray transition energies for elemental analysis\cite{cataldo2022}. However, there has been a growing need in the negative muon community, working on nuclear physics, for software tools that can be used to estimate the 2pF model's $c$ and $t$ parameters from experimental muonic X-ray measurements \cite{Araujo2024, MIZUNO2024, beyer2025, verbeek2023, niikura2024}.

In this work, the aim is to develop \texttt{MuDirac} (1.3.0), so that it can reverse-engineer the Dirac equation to efficiently obtain the 2pF model parameters $c$ and $t$ using, at least, two experimentally measured muonic X-ray transition energies.  \texttt{MuDirac} (1.3.0), is an open, publicly available, sustainable and computationally efficient software tool that will be at the disposal of the negative muon community, so that it can use the code to accurately estimate nuclear properties such as the nuclear charge distribution, $\rho$, and the root-mean-square radius $R_{\textrm{rms}}$. \texttt{MuDirac} will be able to be downloaded from the MSCP's GitHub repository \footnote{https://github.com/muon-spectroscopy-computational-project} and it will be released on the Conda environment \footnote{https://anaconda.org/channels/conda-forge/packages/mudirac/overview}. \newline

\section{2pF model Optimisation Development}

\texttt{MuDirac} calculates muonic x-ray transition energies using the Dirac equation for a radial potential, which can be reduced to a pair of first-order one-dimensional differential
equations\cite{sturniolo2021}:

\begin{align}
     \frac{\partial Q}{\partial r}= \frac{k}{r}Q + 
     \left[ mc- \frac{E-V(r)}{c} \right]P \\
     \frac{\partial P}{\partial r}= -\frac{k}{r}P + 
     \left[ mc+ \frac{E-V(r)}{c} \right]Q 
\end{align}
\label{dirac}

where $P$ and $Q$ represent two components of the equation and $m$ is the mass of the orbiting particle. Atomic units are
used, so that $e = \hbar =1 $ and $c = \frac{1}{\alpha}\approx137$. $k$ is a quantum
number that can take on any positive or negative integer values (but not zero). $V(r)$ is the electrostatic potential generated by
the nuclear charge distribution. Here, $V(r)$ is defined by the 2pF model, and we are looking to optimise the $c$ and $t$ parameters by reverse-engineering equations (\ref{dirac}).  This is: the muonic x-ray transition energies will be taken from experiments\cite{FRICKE1995, megumi2025, Oreskina2017}, and we will calculate $c$ and $t$ in $V(r)$ so that the calculated transition energy reproduces the experimental value. From the optimised values of $c$ and $t$ we build a corresponding 2pF model for nuclear charge distribution, and then use that to calculate the root-mean-square radius $R_{\textrm{rms}}$.  \newline

The definition of the root-mean-square radius $R_{\textrm{rms}}$ is essentially methodological. Experimental values for the muonic x-ray transition energies are fitted by the Dirac equation which, in turn, requires several approximations to be solved. \texttt{MuDirac} can account for the effects of finite nuclear size (via the 2pF model), vacuum polarizability, non-relativistic recoil corrections and electronic screening.  \newline

$R_{\textrm{rms}}$ is defined by equation (\ref{rms}) and equation (\ref{2pf}) defines $\rho(r)$ which, in turn, must fulfil the normalization condition and conserve the total charge:

\begin{equation}
    \label{norm-cond}
    \int \rho(r)4\pi r^{2}\dd r = Ze
\end{equation}

combining (\ref{rms}), (\ref{2pf}) and (\ref{norm-cond}) we can obtain an expression for the mean square radius\cite{Engfer1974}:
    
\begin{multline}
\label{msr}
 R_{rms}^{2} =  \frac{ \int r^{2}\rho(r)\dd r }{\int \rho(r)\dd r} = \\
  = R^{2}\left[ 1- \left[ \exp(-\kappa)\kappa^{-3}6(1+ \kappa^{2}3.029 )\times \left[ 1+ \left(\frac{\pi}{\kappa}\right)^{2} \right]^{-1} \times \left[ 1+ \frac{7}{3}\left(\frac{\pi}{\kappa}\right)^{2} \right]^{-1} \right] + O \exp(-2\kappa) \right]
\end{multline} 

where

\begin{equation}
\label{kappa}
    \kappa = \frac{4\ln(3)c}{t}
\end{equation}

and

\begin{equation}
    \label{R}
    R^{2}=\frac{3}{5}c^{2}\left[ 1+ \frac{7}{3}\left(\frac{\pi}{\kappa}\right)^{2} \right]
\end{equation}

In (\ref{msr}) the term $\exp{-\kappa}\approx 10^{-5}$ for atoms with $ 5 \leq Z$. So, for those atoms, combining (\ref{kappa}) and (\ref{R}), it can be assumed that: 

\begin{equation}
    \label{Rsqr}
    R^{2}=\frac{3}{5}c^2+\frac{7}{5} \left(\frac{\pi t}{4\ln(3)}\right)^{2}
\end{equation}

Hence, for for atoms with $ 5 \leq Z$, the mean square radius $ R_{\textrm{rms}}^{2}\approx R^{2}$ can be obtained using the 2pF analytical description of $\rho(r)$, where the $c$ and $t$ parameters are related by equation (\ref{Rsqr}). It is not possible to use this method for the small set of atoms with $ Z < 5$. For atoms where $ Z < 5$ the term $\exp{-\kappa}$ cannot be considered negligible. Hence, equation (\ref{msr}) cannot be reduced to equation (\ref{Rsqr}), and we cannot assume that $c$ and $t$ are related by equation (\ref{Rsqru}). An integral method is needed to calculate $R_{\textrm{rms}}$ in this case\footnote{We are currently working on this and it will be part of a subsequent publication.}. 

If we define:

\begin{equation}
    \label{a}
    a=\frac{7 \pi^{2}}{48 \ln(3)^{2}}
\end{equation}

and

\begin{equation}
    \label{Ru}
    R_{u} = c^{2} + at^{2}
\end{equation}

Equation (\ref{Rsqr}) can be written as: 

\begin{equation}
    \label{Rsqru}
    R_{\textrm{rms}}^{2}= \frac{3}{5}R^{2}_{u}= \frac{3}{5} (c^{2} + at^{2})
    \end{equation}

Hence, $R_{\textrm{rms}}^{2}$ defines an ellipse in the $c$ and $t$ domain. 

\subsection{Functional relationship between muonic X-ray transition energies and the 2pF model} \label{2pF_func_rel}
\begin{figure}[!htb]
    \centering
    \includegraphics[width=0.8\linewidth]{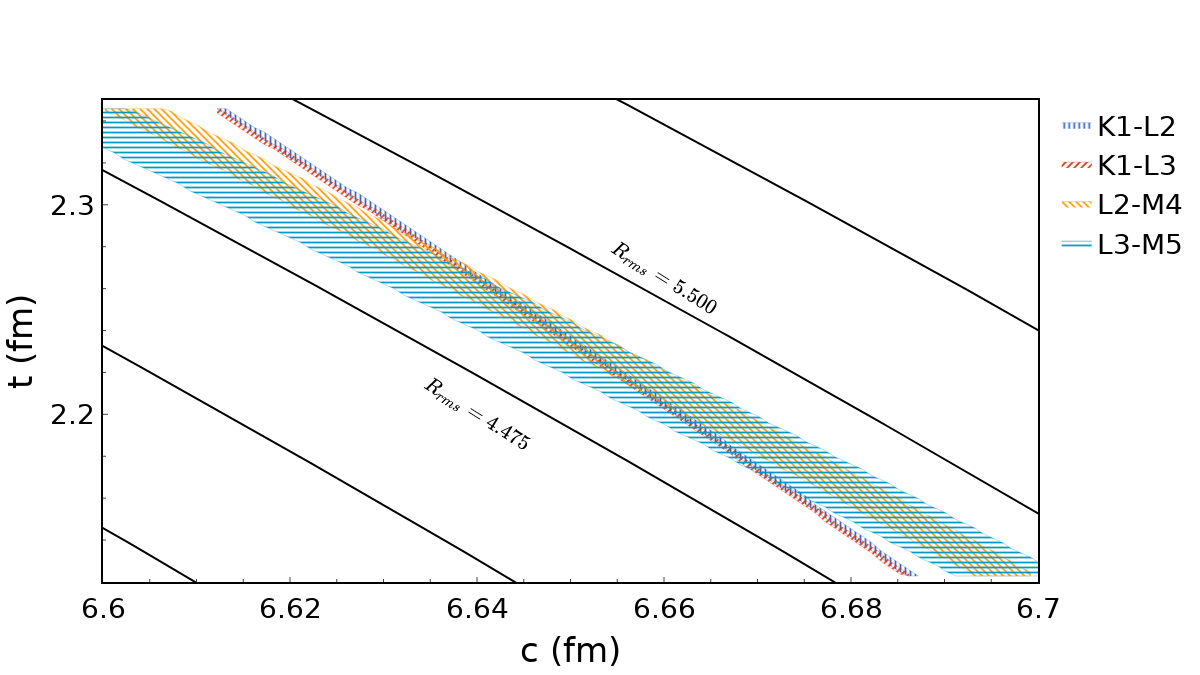}
    \caption{2pF nuclear model parameter bands which reproduce valid muonic X-ray transition energies for four selected transition in $^{206}\text{Pb}$. The width of each band relates to the uncertainty in the experimental measurement of the transition energy.} 
    \label{fig: Pb_206 example bands}
\end{figure}

\vspace{5mm}
The binding energy of the muonic atom is sensitive to the potential of the nucleus, which in
turn is determined by the nuclear charge distribution. In terms of the 2pF model, this means the simulated muonic X-ray transition energies between levels $i$ and $j$, $E_{ij}$, are affected by changes on the $c$ and $t$ parameters of the 2pF model. As we are mainly concerned with muonic X-ray transition energies between energy levels for which we have experimental information, we will represent the transition energies with only one index $k$, which represents a specific energy transition. $E_{ij}=E_{k}$, with, for instance, $k=K1-L2$ or $k=K1-L3$. By reverse-engineering the Dirac equation (\ref{dirac}) for each experimental muonic X-ray transition energy, we can obtain a set of $c$ and $t$ values.  These values are defined by a continuous, functional relationship that can be used to define a ``band'' in the ($c$,$t$) domain, which has a width that is related to the experimental error associated to the muonic X-ray transition energy.  We need at least two of these bands to define a region of the ($c$,$t$) domain where the values of $c$ and $t$ would produce a value of $E_{k}$ that would fall within $E_{k}$'s experimental error. Sometimes, it may be necessary to include more than two bands, as in the case where the lowest $K1-L2$ and $K1-L3$ transitions completely overlap and cannot be distinguished. Figure (\ref{fig: Pb_206 example bands}) is a graphical representation of this method, where we have included higher transitions $L2-M4$ and $L3-M5$. It should be pointed out that this methodology relies on accurate experimental measurements for the transition energies. In particular, care should be taken when when adding higher energy transitions: their experimental errors should be accurately estimated, as their sensitivity to the nuclear charge distribution is smaller\cite{FRICKE1995, megumi2025}.  A similar approach has been recently applied by Saito et. al.\cite{megumi2025}. \newline

In Figure (\ref{fig: Pb_206 example bands}) the black lines are defined by equation (\ref{Ru}), and they represent values of $c$ and $t$ associated to a pre-determined $R_{\textrm{rms}}$ value. Is clear that the bands of feasible Fermi parameters for each transition closely follow the contours of $R_{rms}$. The muonic transition energies are known to be sensitive to $R_{\textrm{rms}}$, however the bands don't coincide with the $R_{\textrm{rms}}$ contours perfectly. This is due to the small model dependence caused by the approximations involved in solving the Dirac equation. Each transition energy has a slightly different model dependence resulting in bands covering different areas of the ($c$,$t$) model parameter space. Hence, there will be a region of the parameter space, $\Omega^*$, where multiple bands overlap. In this region, the nuclear model used will fit the experimental muonic X-ray transition energies used to determine it. This reduces the wide range of valid 2pF parameters to a subset which can be described by a optimal value with small uncertainty determined by a least squares method. The optimal parameters $c^*$ and $t^*$ are found by minimising the difference between the experimental and simulated energies, as defined in equations (\ref{eq: SE}) to (\ref{eq: opt 2pF parameters}):
\begin{align}
SE_{k}(c, t) & = {\frac{(E_{k}^{\exp}-E_{k}^{\textrm{sim}}(c,t))^2}
{\sigma_{k}^2}}, \label{eq: SE}\\[10pt] 
\chi^2(c, t) & = \displaystyle\sum_{k} {SE}_{k}(c, t) \label{eq: chi sq}, \\[10pt] 
\Omega^* & = \{ c, t \in \Omega : \forall k \in I,\; SE_{k}(c,t) < 1 \}, \label{eq: 2pF valid domain}\\[10pt] 
(c^*, t^*) & = \arg\min_{c,t \in \Omega}\chi^2(c,t) \label{eq: opt 2pF parameters},\\[10pt] \notag{}
\end{align}
where $I$ define the space containing the indexes $k$, of the selected transitions. $E_{k}$ is the energy of a selected transition between two given states, and $\sigma_{k}$ is the associated experimental uncertainty. $\chi^{2}(c,t)$ represents the sum of all the differences between simulated and experimental energies for all the energy transitions considered in a muonic atom, and $(c^{*},t^{*})$ is the point in the subset of the $(c,t)$ domain where $\chi^{2}(c,t)$ reaches a minimum. \newline

In the literature that uses this method, the overlapped region $\Omega^*$ is found by brute force \cite{megumi2025}. The entire domain of the 2pF parameters is scanned within reasonable search bounds to find $c^*$ and $t^*$. Often, only the $K1-L2$ and $K1-L3$ transitions are measured and their 
associated bands coincide completely,
as seen in figure (\ref{fig: Pb_206 example bands}). This means the feasible set cannot be reduced. In other cases, only the $K1-L2$ transition has been measured. In these cases, the strategy with a few of these transitions has been to fix $t=\qty{2.3}{fm}$ and use $c$ as a fitting parameter\cite{Engfer1974}. 

\subsection{Optimisation Approach}\label{sec: 2pF opt dev}
\vspace{5mm}

As seen in subsection (\ref{2pF_func_rel}), the optimal Fermi parameters have been found using a brute force least squares fitting method that involves iterating over the ($c,t$) space until a reasonably accurate $(c^*, t^*)$ pair is found. This method is inherently wasteful of computational resources and, in the next subsections, we will show how the parameter search can be optimised.   This is achieved in two parts: by better parametrising the 2pF model coordinate system, and by using more efficient numerical methods to solve the least squares problem. These two parts are described in the following sub-sections.

\subsubsection{Choice of coordinate system} \label{sec: coord choice}
\vspace{5mm}

As the valid bands for c and t are generally parallel to the $R_{\textrm{rms}}$ contours in figure (\ref{fig: Pb_206 example bands}), we choose to parametrise the Fermi parameters using polar coordinates. Doing so makes the brute force method for calculation more efficient by removing points far from the relevant $R_{\textrm{rms}}$ contour from consideration. Effectively, we create more restricted bounds for scanning of the $\Omega^{*}$ domain. \newline 

Equation (\ref{Ru}),  describe an origin centred ellipse in the ($c$,$t$) space, where $R_{\textrm{rms}}^2$ is the semi major axis and $\frac{R_{\textrm{rms}}^2}{a}$ is the semi minor axis. Parametrising the ellipse equation, we get the following equations in polar coordinates $(R_{\textrm{rms}}, \theta)$:

\begin{align}
\quad 0 \leq \theta \leq \frac{\pi}{2},& \quad R^{\min}_{\textrm{rms}} \leq R_{\textrm{rms}} \leq R^{\max}_{\textrm{rms}},\label{eq: polar bounds}\\[15pt]
 r(\theta, R_{\textrm{rms}})  &= \frac{\sqrt{\frac{5}{3}}R_{\textrm{rms}}}{\sqrt{\cos^2\theta+a \cdot \sin^2\theta}},\label{eq: polar r}\\[15pt]
c  &= r(\theta, R_{\textrm{rms}})\cos\theta, \label{eq: polar c}\\[8pt]
t  &= r(\theta, R_{\textrm{rms}}) \sin\theta.\label{eq: polar t}\\[8pt] \notag{}
\end{align}

In equation (\ref{eq: polar bounds}), $R^{\min}_{\textrm{rms}}$ and $R^{\max}_{\textrm{rms}}$ are defined using equation (\ref{Rsqru}).
Using this polar coordinate system, we were able to scan the Fermi parameters along contours of $R_{\textrm{rms}}$ by varying $\theta$. To convert back to the conventional $c$ and $t$ coordinates we use equations (\ref{eq: polar c}) and (\ref{eq: polar t}). The use of polar coordinates is computationally more efficient than using brute force methods, and we will show that it is also beneficial for the optimization algorithms that we are using. \newline 

\begin{figure}[!htb]
        \centering
        \includegraphics[width=0.9\textwidth]{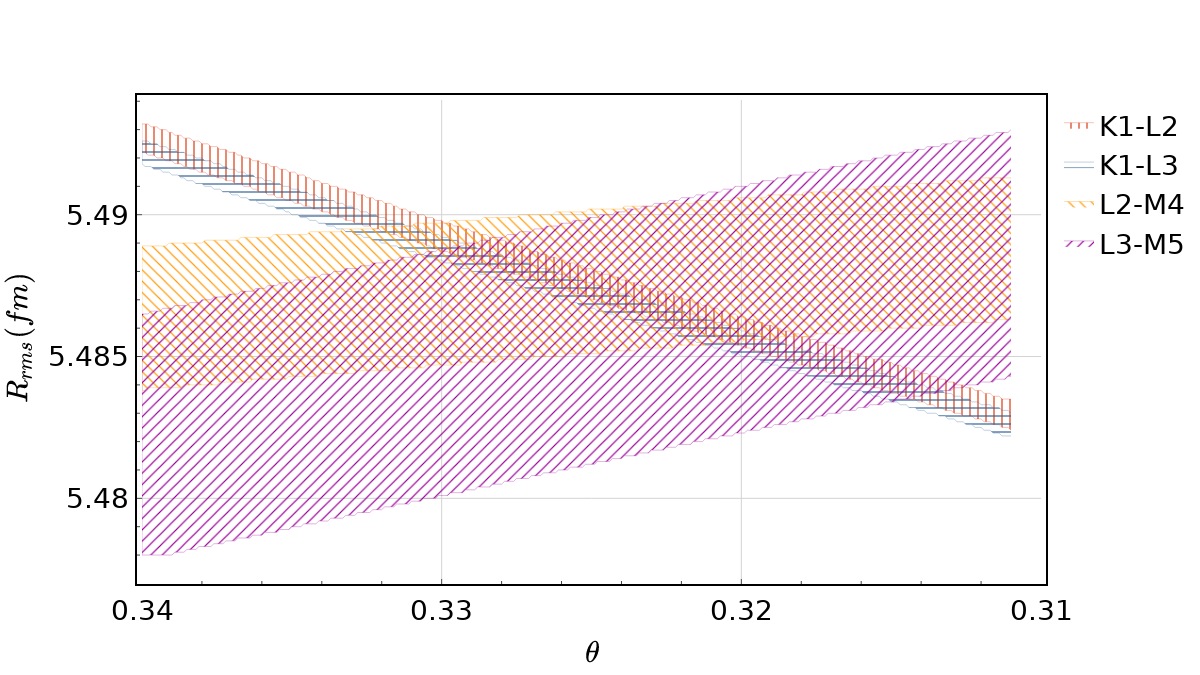}
        \caption{2pF nuclear model parameter bands, which reproduce valid muonic X-ray energies for four selected transition in $^{206}\text{Pb}$, now shown in polar coordinates. The coordinate system better illustrates the slight model dependence of the $R_{\textrm{rms}}$ associated with each band. This means that the model solutions for each band do not perfectly coincide, causing a more restricted overlap region for the model parameters that reproduce the experimental results for all the bands.\label{fig: polar bands}}
\end{figure}

Figure (\ref{fig: polar bands}) shows the same bands of feasible 2pF parameters in $^{206}\text{Pb}$ as shown in figure (\ref{fig: Pb_206 example bands}), only in polar coordinates. $\theta$ is reversed in the x axis to follow the shape of the bands in figure (\ref{fig: Pb_206 example bands}). This effect arises due to how theta is defined in the parametrisation. The figure shows more clearly how the feasible bands of polar coordinates are dependent on $R_{\textrm{rms}}$ and how using this coordinate system makes it easier to define a square domain for the search. This is beneficial for scanning the domain as it avoids scanning regions of the ($c$,$t$) space which are far from a relevant $R_{\textrm{rms}}$ contour. This is also beneficial for the efficiency of the numerical methods, as we will show in the next subsection. \newline 

\begin{figure}[!htb]
        \includegraphics[width=0.9\textwidth]{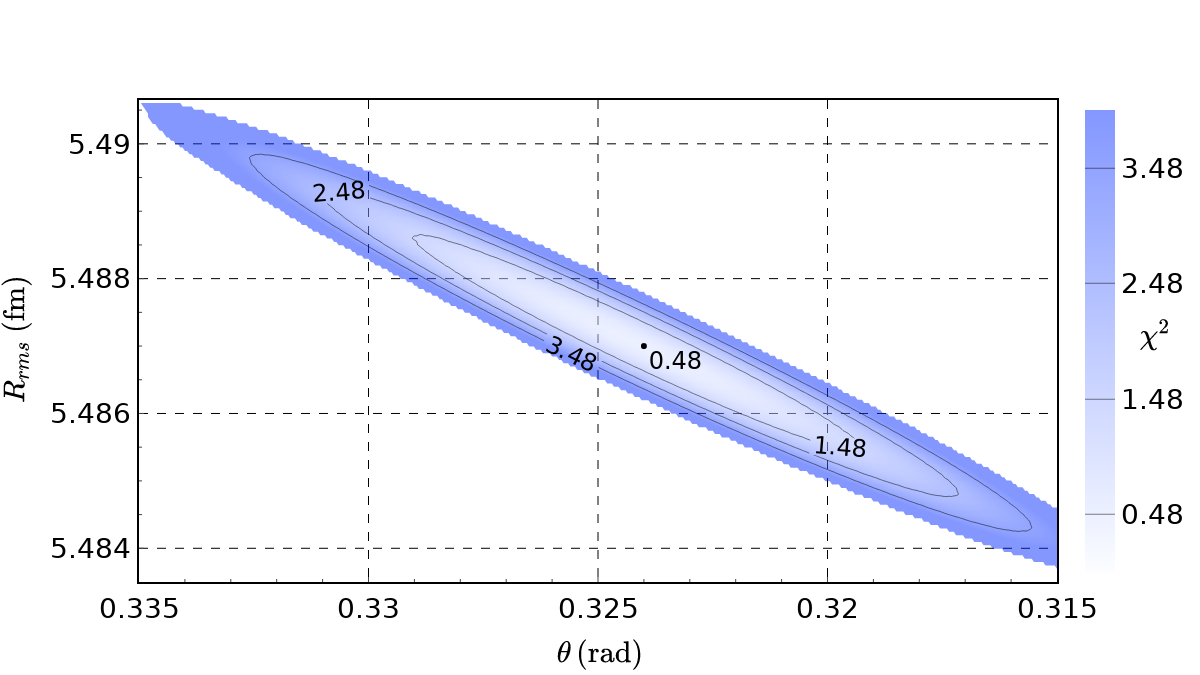}
        \caption{$\chi^2$ contours for $^{206}\text{Pb}$ in polar coordinates. $\chi^2$ is calculated using the experimental energies and the \texttt{MuDirac} simulated energies, for the same 4 transitions,  over a polar coordinate grid with a resolution of 4 decimal places. The contours mark the $\chi^2$ minimiser and lines enclosing regions of $\chi_{\min}^2 +n$ for each $n\sigma$ level of uncertainty. The $\chi^2_{\min}=0.48$ is found at $R_{\textrm{rms}}=5.487, \; \theta=0.324$.\label{fig: chi sq heat map} }
\end{figure}

Figure (\ref{fig: chi sq heat map}) was generated by calculating $\chi^2$ against experimental data obtained from\cite{Engfer1974}. The contours showing the uncertainty regions are elliptical and narrow in $R_{\textrm{rms}}$, demonstrating how $R_{\textrm{rms}}$ can be extracted to good precision despite the uncertainties in $c$ and $t$. The wider uncertainty in $\theta$ encapsulates the uncertainty arising from the model dependence. \newline 


 To verify that describing the 2pF parameters in polar coordinates improves the efficiency for the reverse-engineering of the Dirac equation, we need to benchmark the results that use polar parametrisation. In the next subsection the choice of numerical optimisers is discussed and the benchmarking process is presented.
 \newline

\subsubsection{Optimisation Algorithms}
\label{sec: opt algos}
The new coordinate system significantly improves the brute force method. However, the region of valid Fermi parameters, if it exists, is small and extremely sensitive to small changes. This means that an effective optimisation method needs scans of the 2pF domain to more than 4 decimal places. This requires very carefully selected bounds for the search domain of each atom. We therefore decide that optimization routines with minimal user input are a better choice. Equations (\ref{eq: chi sq}) and (\ref{eq: opt 2pF parameters}) describe the unconstrained minimisation problem without loss of generality with respect to the coordinate system. Properties of continuity and differentiability are empirically assumed. \newline 

There are many numerical algorithms which can be used to solve this least squares problem. Choosing the most efficient one involves programming many minimisers into \texttt{MuDirac} and comparing them across a range of datasets. For this comparison, we used the FitBenchmarking\cite{Dolan2002-cz} tool, to benchmark a range of commonly used minimisers for the \texttt{MuDirac} problem. 

\subsubsection{Evaluating optimisation algorithms with FitBenchmarking}
 FitBenchmarking solves optimisation problems using a variety of commonly used fitting algorithms in Python and provides evaluations based on improving the accuracy of the minimisation and the amount of runtime \cite{Dolan2002-cz}. The results of this first benchmarking allow us to perform a first selection of potential optimisers to include in \texttt{MuDirac}. \newline



\begin{figure}[!htb]
    \centering
\includegraphics[width=0.7\linewidth]{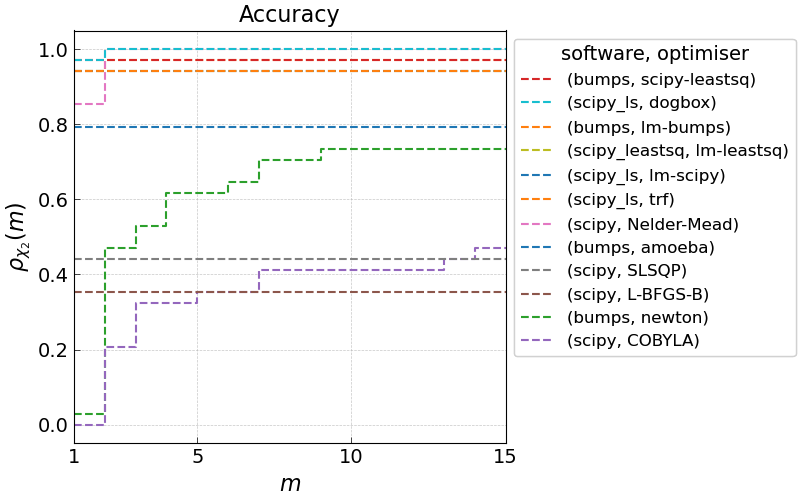}
    \caption{Optimisation accuracy of different optimisation algorithms in \texttt{MuDirac} using 24 muonic atom datasets analysed with FitBenchmarking. 
    The axes are defined in equation (\ref{eqn: optimisation1}).
    The x axis shows $m$ multiples of the $\chi^2_{\min}$ for the datasets. The y axis gives the proportion of datasets for which each optimiser found a minimum within $m$ multiples of the $\chi^2_{\min}$. The best optimisers are found in the top left region of the plot. The legends are shown in order of descending $\rho_{\chi^2}(n=1)$. Optimisation algorithms which failed to converge have been omitted from the plot for clarity.} 
    \label{fig: fitbenchmarking accuracy}
\end{figure}

\begin{figure}[!htb]
    \centering
    \includegraphics[width=0.7\linewidth]{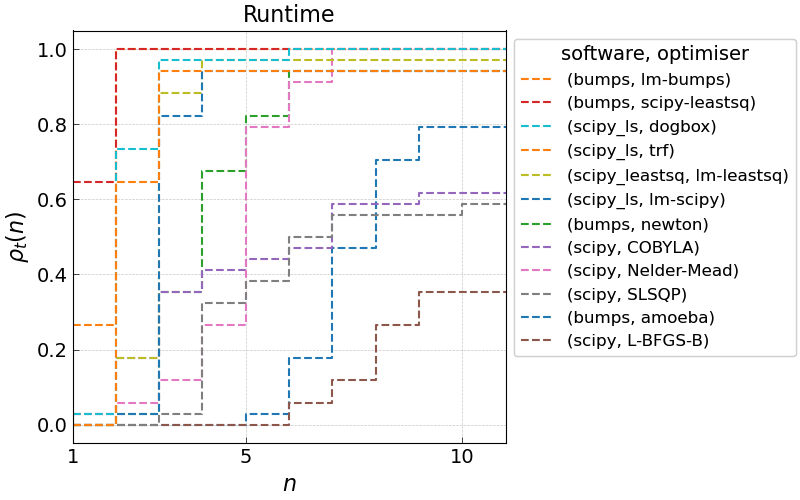}
    \caption{Optimisation runtime of different optimisation algorithms in \texttt{MuDirac} using 24 muonic atom datasets analysed with FitBenchmarking. The axes are defined in equation (\ref{eqn: optimisation4})The x axis gives the runtime $t$ rescaled to multiples, $n$, of $t_{min}$ for each dataset. The y axis gives the proportion of datasets for which each optimiser found a minimum within that multiple of the $t_{\min}$ from all the optimisers. The best optimisers are found in the top left region of the plot. The legend is shown in order of descending $\rho_{t}(n=2)$. Optimisation algorithms which failed to converge have been omitted from the plot for clarity.}
    \label{fig: fitbenchmarking time}
\end{figure}

The figures (\ref{fig: fitbenchmarking accuracy}) and (\ref{fig: fitbenchmarking time}) are used to determine the best optimisation algorithms to use in \texttt{MuDirac}.  For any optimisation problem, the requirement is to provide an accurate solution by minimising the computational costs across all possible datasets. \newline 

For a given muonic atom, the variable that is optimised when looking for optimum values $c^{*}$ and $t^{*}$ is $\chi^2(c,t)$, as defined by equation (\ref{eq: chi sq}).  For determining the best optimisation algorithms, we optimised $\chi^2(c,t)$ in 24 muonic atoms datasets taken from Engfer et. al\cite{Engfer1974}\footnote{The full 24 atoms dataset is given in the Supplementary Information. There were accurate experimental measurements of the transition energies available for these atoms.}. This is, we run all the available optimisation algorithms on all the 24 datasets.  After scanning all 24 datasets, using all the available optimisation algorithms, we can define a minimum $\chi^2_{\min}(c,t)$. The figure axes are then defined by the following equations: 

 \begin{align}
    \chi^2_{\min}(x_i) &= \min_{s \in S}(\chi^2_s(x_i)) \label{eqn: optimisation1} \\
    t_{\min}(x_i) &= \min_{s \in S}(t_s(x_i)) \label{eqn: optimisation2} \\
    \rho_{\chi^2}(m) &=\mathbb{P} \left( \chi^2 (x_i)\leq m \; \chi^2_{\min}(x_i) | x_i \in X \right)\label{eqn: optimisation3} \\
    \rho_{t}(n) &=\mathbb {P} \left( t (x_i)\leq n \; t_{\min}(x_i) | x_i \in X \right) \label{eqn: optimisation4}
 \end{align}

where $S$ is the set of optimisers, $X$ is the set of muonic atom transition datasets and $n, m$ are natural numbers. For each muonic atom dataset $x_i \in X $ there will be a particular optimisation algorithm $s \in S$ that will provide the smallest $\chi^2_{\min}(c,t)$. The x-axis of Figure (\ref{fig: fitbenchmarking accuracy}) represents multiples of these $\chi^2_{\min}(c,t)$, while the y-axis represents the proportion of datasets for which  a given optimiser found a value of $\chi^2(c,t)$ that satisfies $\chi^2(c,t) \leq \chi^2_{\min}(c,t)$. 

Hence, the best optimisers in terms of  accuracy and are found in the upper left region of Figure (\ref{fig: fitbenchmarking accuracy}). Note that because $\rho$ is calculated as a comparison to the best optimiser, the relative order of the following minimisers does not represent their performance relative to each other. If the best optimiser was removed from the analysis the results for the rest of the optimisers would change and may be ordered differently.\newline 

Similarly, Figure (\ref{fig: fitbenchmarking time}) shows the best optimisers in terms of running time, i.e.: which ones are the fastest ones.  After scanning all 24 datasets, using all the available optimisation algorithms, we can define a best optimisation time for optimising $\chi^2(c,t)$.  There will be a particular optimisation algorithm, applied to a particular muonic atom, that will be the fastest in reaching an optimised $\chi^2(c,t)$ value. The x-axis in Figure (\ref{fig: fitbenchmarking time}) represent multiples of the minimum time required for optimising $\chi^2(c,t)$. Hence, the best optimisers in terms of  accuracy and are found in the upper left region of Figure (\ref{fig: fitbenchmarking time}). 
\newline

 The x-axes in Figures (\ref{fig: fitbenchmarking accuracy}) and (\ref{fig: fitbenchmarking time}) indicate multiples of the best possible accuracies or runtimes found by the set of optimisation algorithms. This allows the algorithms to be compared across datasets. And it can be seen that, on the one hand, the best possible accuracy for all the datasets matched those found using brute force, on the other hand, the best possible runtime was on the order of a minute or so compared to the hours taken for high resolution by brute force. \newline

In our analysis the best algorithms were the Scipy least squares (scipy-leastsq) algorithms \cite{2020SciPy-NMeth}. These algorithms use the C++ Minpack libraries, and we have implemented the Ceres solver \cite{Agarwal_Ceres_Solver_2022} for \texttt{MuDirac}. Ceres solver is an open-source C++ library designed for solving optimization problems, which is recommended by the C++ Minpack developers. The Ceres solver comes with range of optimization algorithms similar to Scipy which is used for initial analysis.   \newline

\subsubsection{Benchmarking optimisation methods in \texttt{MuDirac}}
\label{sec: md benchmarking}

\begin{figure}[!htb]
    \centering
    \includegraphics[width=0.9\linewidth]{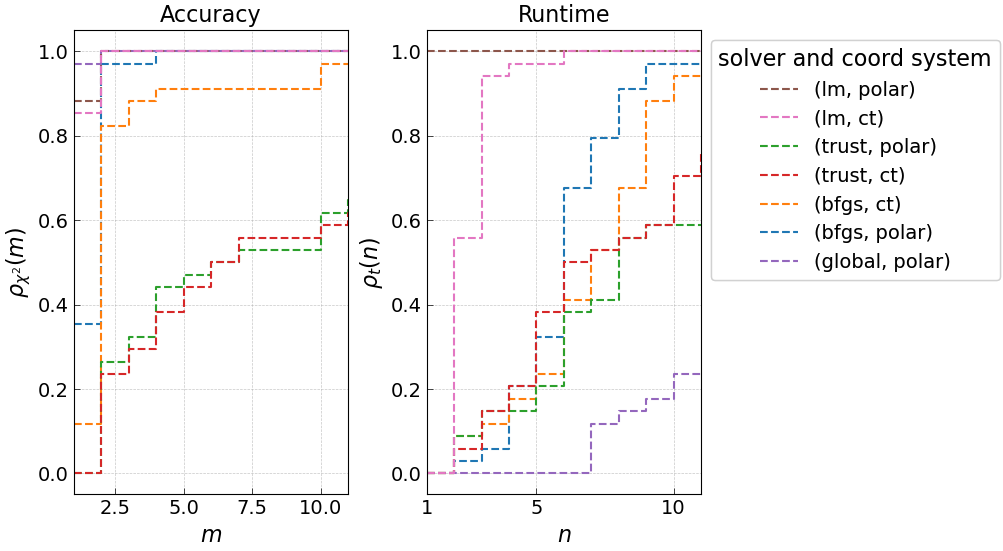}
    \caption{Optimisation accuracy and runtime of \texttt{MuDirac} using different optimisation algorithms and coordinate systems using 24 muonic atom datasets. The axes of the plots follow \cref{fig: fitbenchmarking accuracy,fig: fitbenchmarking time} using equations (\ref{eqn: optimisation3}) and (\ref{eqn: optimisation4}). The accuracy plot (left) shows the true $\chi^2_{\min}$ is found most frequently by the global optimiser as expected, shortly followed by the lm optimisers. The runtime plot (right) shows that the lm optimiser using polar coordinates is the fastest. Together the plots show that the most reliable and efficient solver is the lm solver using the polar coordinate system.} 
    \label{fig: opt accuracy and runtime}
\end{figure}

The first FitBenchmarking analysis has allowed us to select a reduced set of optimisation algorithms to implement in \texttt{MuDirac}, which includes the Levenberg-Marquardt (lm)\cite{lm_algo} and the Broyden–Fletcher–Goldfarb–Shanno (BFGS) algorithms. These optimisers efficiently scanned the $(c,t)$ domain to find local minima for $\chi^2(c,t)$. To increase the robustness of the accuracy comparison, we used a global optimiser that took large enough runtime to scan the whole $(c,t)$ domain, in cartesian and polar coordinates, and look for the best optimisation algorithms.  Figure (\ref{fig: opt accuracy and runtime}) show the analysis for accuracy and runtime efficiency for the best optimisation algorithms evaluated using the 24 muonic atom datasets\cite{Engfer1974} used in Figures (\ref{fig: fitbenchmarking accuracy}) and (\ref{fig: fitbenchmarking time}). The polar coordinate system is that described in section (\ref{sec: coord choice}).

Figure (\ref{fig: opt accuracy and runtime}) shows the Levenberg-Marquardt (lm) method in polar coordinates as the most efficient and accurate. The figures also show that using polar coordinates is generally beneficial across all optimisers, and that the BFGS method performed less efficiently than the lm algorithm. For our purposes, is sufficient to only include the lm algorithm and polar coordinate system in \texttt{MuDirac} for the optimisation of the 2pF parameters. \newline

The new software details for using the intended new functionality in MuDirac are described in the next section.

\section{Software Implementation details}
\label{sec: soft implementation}
Following the benchmarking, we implemented
the lm algorithm, run on a polar coordinate system,  in \texttt{MuDirac} (1.3.0) for the optimisation of the 2pF parameters. \newline 

All the documentation for the current version of \texttt{MuDirac} is in the Muon Spectroscopy Computational Project (MSCP)'s website\footnote{https://muon-spectroscopy-computational-project.github.io/mudirac/index.html}. For running the optimisation of the 2pF model's parameters, new keywords have been added to \texttt{MuDirac}'s main input file, \texttt{<name>.in}, and an additional input file, \texttt{<name>.xr.in}, has been created for providing the muonic x-rays experimental measurements. Examples of the new keywords needed for the main input file are shown in \textcolor{red}{red} in the  $^{197}\text{Au}$ example below.

\begin{Verbatim}[commandchars=\\\{\}]
#  Au 197 input file 1 example; <name>.in 
\textbf{element}: Au
isotope: 197
xr_lines: K1-L2,K1-L3,L2-M4,L3-M5
uehling_correction: True
reduced_mass: True
electronic_config: [Au]
\textcolor{red}{nuclear_model: FERMI2}
\textcolor{red}{fermi_c: 6.6}
\textcolor{red}{fermi_t: 2.0}
\textcolor{red}{optimise_fermi_parameters: True}
\textcolor{red}{rms_radius_decimals: 4}

\end{Verbatim}

In this input file the optimisation is turned on the by \texttt{optimise$\_$fermi$\_$parameters: True} keyword.
The Fermi model for the nuclei with variable $c$ and $t$ is chosen by the \verb|nuclear_model: FERMI2| parameter. The Fermi parameters provided are used as the starting point for the optimisation. If they are not provided the optimisation will start from the default values from nuclear data tables\cite{marinova_2013}. The \texttt{rms$\_$radius$\_$decimals: 4} keyword controls the print precision of the optimised values. \newline 

For the second input file, a placeholder text file is being used temporarily to contain the x-ray measurement data, as there is currently no file standard for this experimental data. Muonic x-ray energies and their corresponding experimental uncertainties should be provided in eV. An example file for $^{197}\text{Au}$ is shown below\cite{66AB1}.

\begin{Verbatim}[commandchars=\\\{\}]
#  Au 197 input file 2 example <name>.xr.in 
\end{Verbatim}
\begin{center}
\begin{tabular}{lllll}
xr$\_$lines: & $K1-L2$ & $K1-L3$ & $L2-M4$ & $L3-M5$ \\
xr$\_$energy: & 5592800.0 & 5762500 & 2474400 & 2343100.0 \\
xr$\_$error: & 5000.0 & 5000.0 & 2000.0 & 2500.0 \\
\end{tabular}
\end{center}

Note that the keyword  \verb|xr_lines| appears in both input files, and its value must be the same in both files. This is because \texttt{MuDirac} must simulate the same transitions contained in the file with the experimental data.  \newline 

Finally to enable optimisation, a separate installation of Ceres solver is required \cite{Agarwal_Ceres_Solver_2022}. Ceres is a  large software compared to \texttt{MuDirac}. But only a minimal installation of Ceres is required for our purposes.  \newline

Instructions on how to install \texttt{MuDirac} are in the MSCP's website. From the command line, \texttt{MuDirac} (1.3.0) should be run with both input files as arguments as follows:
\begin{verbatim}
$\path\to\mudirac <name>.in <name>.xr.in
\end{verbatim}

An additional output file , \verb|<name>fermi_parameters.out| is generated when optimisation is turned on. The file contains the 2pF parameters in both coordinate systems and the $\chi^2$. The result from the $^{197}\text{Au}$ example is shown below:

\begin{Verbatim}[commandchars=\\\{\}]
# Z = 79, A = 197 amu, m = 206.768 au
\end{Verbatim}
\begin{center}
\begin{tabular}{llllll}
fermi$\_$c	& fermi$\_$t &	rms$\_$radius [fm] &	theta &	chi$\_$sq &	time [s] \\
6.6744 &	1.9265 & 5.4207 & 0.2810 & 0.1560 & 59.1980
\end{tabular}
\end{center}

By using the Ceres solver, it is easy to add and compare more least squares minimisation algorithms. Extending the functionality to more nuclear models is also straight forward. The performance of the current implementation for the 2pF model is discussed in the next section.

\section{Results}  
\vspace{5mm}

\begin{figure}[!htb]
    \centering
    \includegraphics[width=0.7\linewidth]{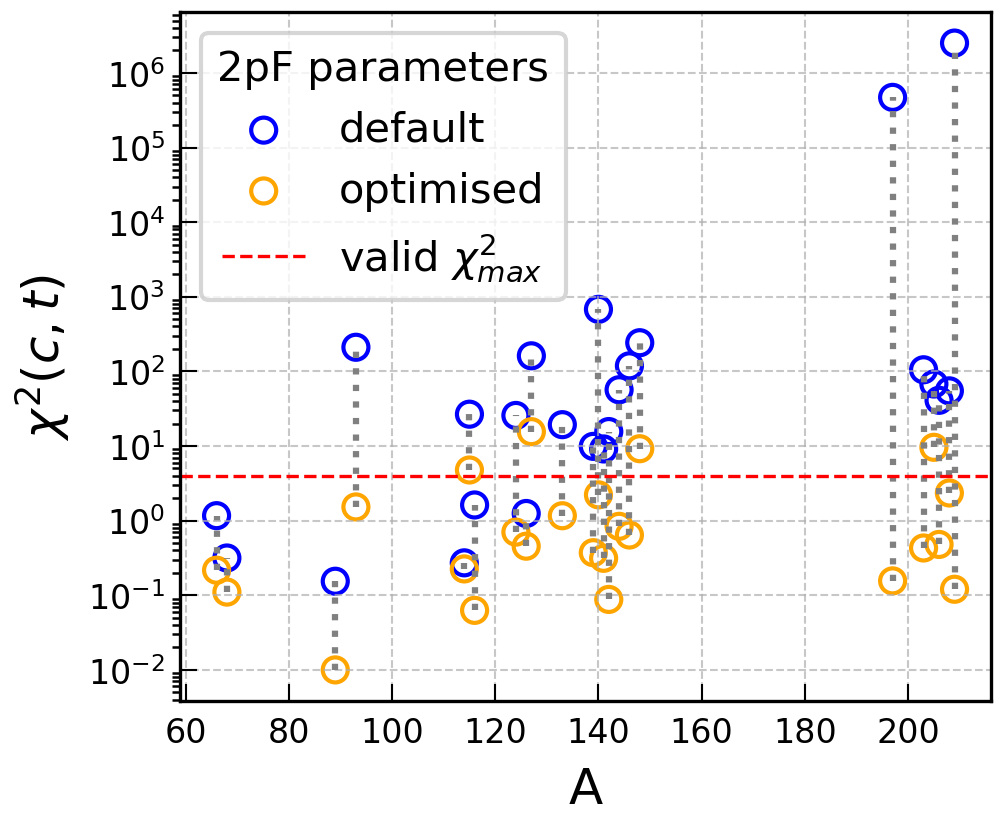}
    \caption{$\chi^2(c,t)$ for a subset of 24 muonic atoms \cite{Engfer1974}. The $\chi^2(c,t)$ is calculated across 4 transitions for each atom using the default 2pF values (blue) and fitted 2pF values (orange) in \texttt{MuDirac}. The horizontal red dashed line at $\chi^2(c,t) = 4$ marks the $\chi^2(c,t)_{\max}$ where the simulated energies can still be within the uncertainty of the measured energies. For clarity of the plot, data from separate sources on the same isotopes have been omitted, with a preference for recency. }
    \label{fig: log chi sq results}
\end{figure}

Figure (\ref{fig: log chi sq results}) presents values of $\chi^2(c,t)$ for a subset of 24 muonic atoms.  The experimental uncertainties for these atoms were taken from Engfer et. al\cite{Engfer1974}. $\chi^2(c,t)$ was calculated using four energy transitions for each atom using the current version of \texttt{MuDirac} (1.2.4) (blue) -where the 2pF model uses $t=\qty{2.3}{fm}$ and $c$ is obtained from nuclear data tables\cite{marinova_2013}- and the new version of \texttt{MuDirac} (1.3.0), using fitted 2pF values (orange). The horizontal red dashed line at $\chi^2(c,t)$=4 marks where the simulated energies can still be within the uncertainty of the measured energies. It can be clearly seen that \texttt{MuDirac}(1.3.0) was able to improve the precision of the simulations and reduce $\chi^{2}(c,t)$ in all cases. \newline 

\begin{figure}[!htb]
    \centering
    \includegraphics[width=0.7\linewidth]{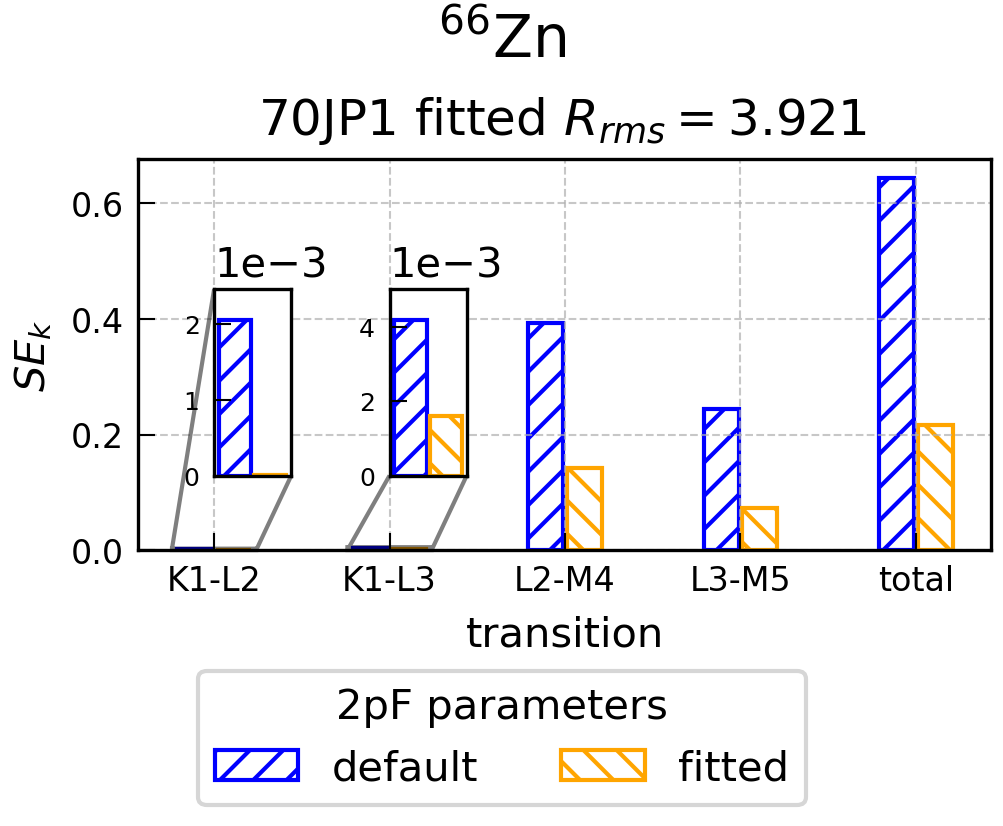}
    \caption{The 70JP1 label at the top refers to the reference for the experimental data in Engfer et. al\cite{Engfer1974}. $\chi^2(c,t)$ and $SE_{k}(c,t)$ calculated at four transitions in $^{66}\text{Zn}$\cite{70JP1}. The first four pair of bars from the left show the individual energy transitions $SE_{k}(c,t)$, separated for each transition, while the last pair shows the $\chi^2(c,t)$.  The bars show the default (blue) and fitted (orange) 2pF values. The  $SE_{k}(c,t) < 1$ and $\chi^2(c,t) < 1$ for all the transitions using the default parameters, meaning the default parameters are in the feasible set. The $SE_{k}(c,t)$ are improved using the fitted parameters for all the transitions. Results for $^{66}\text{Zn}$ are show as an example of the improvement using fitting in MuDirac for light atoms.}  
    \label{fig: opt Zn 66 chi bars}
\end{figure}

\begin{figure}[!htb]
    \centering
    \includegraphics[width=0.7\linewidth]{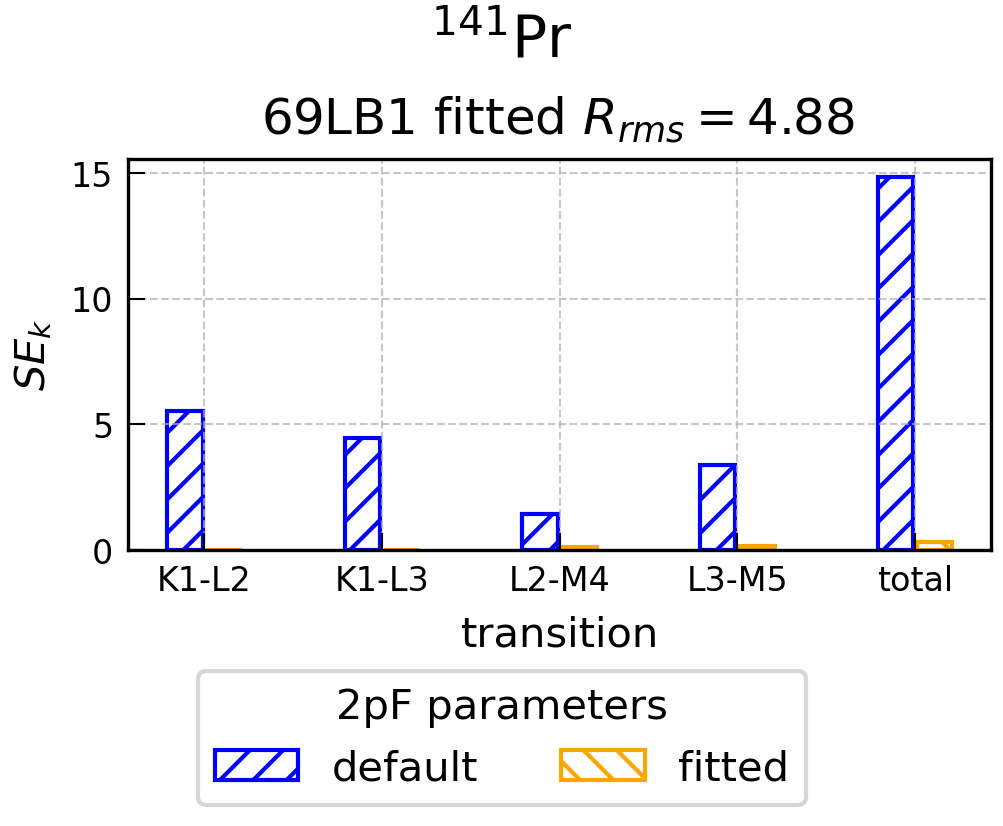}
    \caption{The 69LB1 label at the top refers to the reference for the experimental data in Engfer et. al\cite{Engfer1974}. $\chi^2(c,t)$ and $SE_{k}(c,t)$ calculated at four transitions in $^{141}\text{Pr}$ \cite{69LB1}. The first four pair of bars from the left show the individual energy transitions $SE_{k}(c,t)$, separated for each transition, while the last pair shows the $\chi^2(c,t)$.  The bars show the default (blue) and fitted (orange) 2pF values. The  $SE_{k}(c,t) < 1$ and $\chi^2(c,t) < 1$ for all the transitions using the default parameters, meaning the default parameters are in the feasible set. The $SE_{k}(c,t)$ are improved using the fitted parameters for all the transitions. Results for $^{141}\text{Pr}$ are show as an example of the improvement using fitting in MuDirac for medium sized atoms. }  
    \label{fig: opt Pr 141 chi bars}
\end{figure}

\begin{figure}[!htb]
    \centering
    \includegraphics[width=0.7\linewidth]{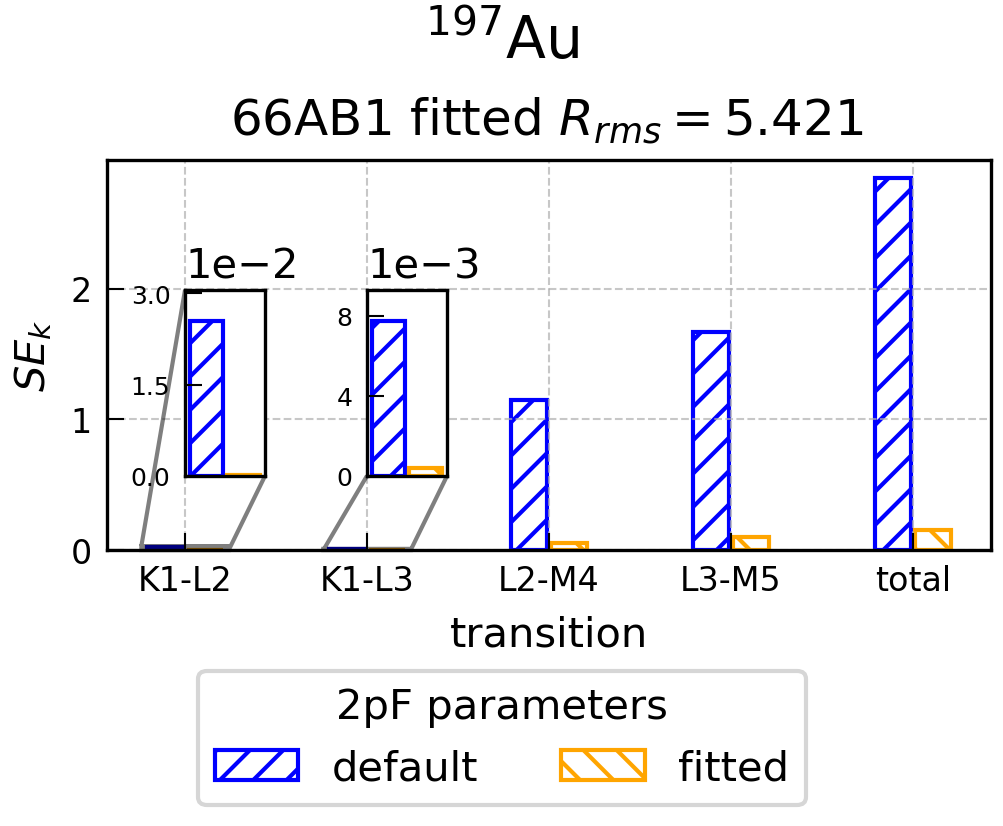}
    \caption{The 66AB1 label at the top refers to the reference for the experimental data in Engfer et. al\cite{Engfer1974} $\chi^2(c,t)$ and $SE_{k}(c,t)$ calculated at four transitions in $^{197}\text{Au}$ \cite{66AB1}. The first four pair of bars from the left show the individual energy transitions $SE_{k}(c,t)$, separated for each transition, while the last pair shows the $\chi^2(c,t)$.  The bars show the default (blue) and fitted (orange) 2pF values. The  $SE_{k}(c,t) < 1$ and $\chi^2(c,t) < 1$ for all the transitions using the default parameters, meaning the default parameters are in the feasible set. The $SE_{k}(c,t)$ are improved using the fitted parameters for all the transitions. Results for $^{197}\text{Au}$ are shown as an example of the improvement using fitting in MuDirac for heavy atoms.}  
    \label{fig: opt Au 197 chi bars}
\end{figure}

\begin{figure}[!htb]
    \centering
    \includegraphics[width=1.0\linewidth]{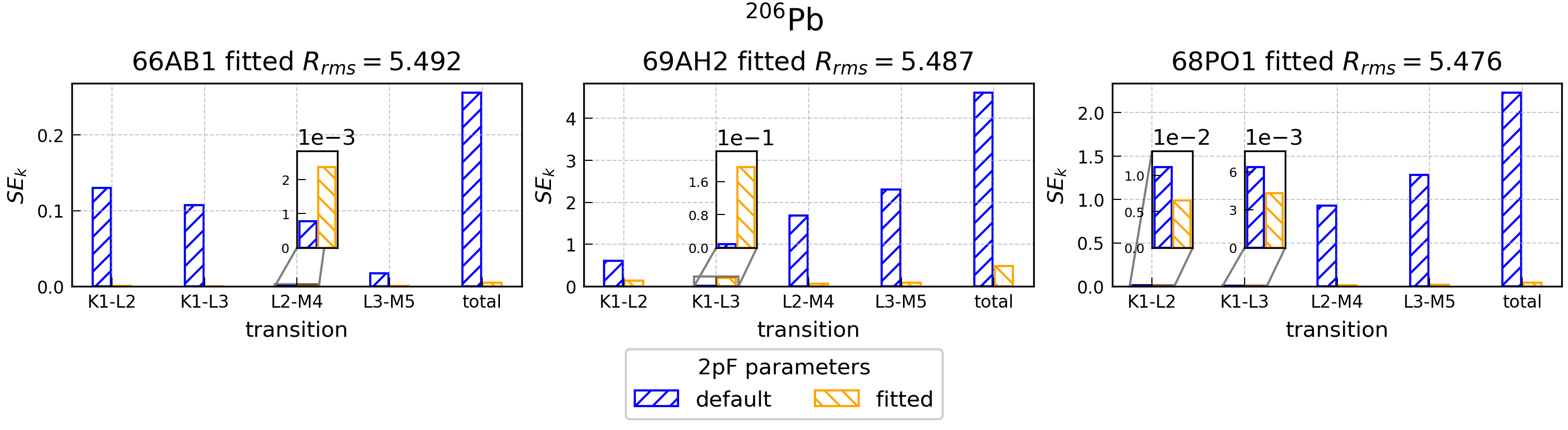}
    \caption{The 66AB1, 69AH2 and 68PO1 labels at the top refer to the reference for the experimental data in Engfer et. al\cite{Engfer1974}. $\chi^2(c,t)$ and  $SE_{k}(c,t)$ calculated at four transitions in $^{206}\text{Pb}$, with the experimental data taken from three different experiments \cite{66AB1, 68PO1, 69AH2}. The first four pair of bars from the left show the individual energy transitions $SE_{k}(c,t)$, separated for each transition, while the last pair shows the $\chi^2(c,t)$.  The bars show the default (blue) and fitted (orange) 2pF values. The  $SE_{k}(c,t) < 1$ and $\chi^2(c,t) < 1$ for all the transitions using the default parameters, meaning the default parameters are in the feasible set. Although the $SE_{k}(c,t)$ are not improved in two individual energy transitions (L2-N4 in 66AB1, K1-L3 in 69AH2), the $\chi^2(c,t)$ are improved using the fitted parameters in the three experiments.  These results for $^{206}\text{Pb}$ are shown as an example of how the 2pF parameters can be improved  across different experimental results. }  
    \label{fig: opt Pb 206 chi bars}
\end{figure}

Figures (\ref{fig: opt Zn 66 chi bars}), (\ref{fig: opt Pr 141 chi bars}), (\ref{fig: opt Au 197 chi bars}) and (\ref{fig: opt Pb 206 chi bars}) show how the 2pF fitted parameters improve the differences between simulated and experimental energies, $\chi^2(c,t)$ and $SE_{k}(c,t)$, for specific muonic x-ray transitions in representatives of light, $^{66}\text{Zn}$, medium, $^{141}\text{Pr}$, and heavy atoms, $^{197}\text{Au}$. Figure (\ref{fig: opt Pb 206 chi bars}) shows how the 2pF fitted parameters improve the differences between simulated and experimental energies, $\chi^2(c,t)$ and $SE_{k}(c,t)$, in $^{206}\text{Pb}$. The experimental transition energies for $^{206}\text{Pb}$ come from three different sources\cite{66AB1, 68PO1, 69AH2}. It can be clearly seen that \texttt{MuDirac}(1.3.0) was able to improve the precision of the simulations, reducing $\chi^{2}(c,t)$ but also the difference between experiment and simulation for individual energy transitions, $SE_{k}(c,t)$, in all cases. \newline

Table (\ref{tableRrms}) shows \texttt{MuDirac} results for $R_{rms}$ for $\mathrm{^{89}Y}$, $\mathrm{^{114}Cd}$, $\mathrm{^{197}Au}$ and $\mathrm{^{206}Pb}$ and compares them to reference results from Engfer\cite{Engfer1974}. \texttt{MuDirac} results agree very well with the reference, experimental results. It is important to note that the uncertainties included do not represent the ground truth uncertainties in the parameters. Instead, they describe the uncertainty conditioned on accurate MuDirac energy simulations. Currently the uncertainties can be estimated graphically in a similar way to the previously described brute force methods, except on a smaller parameter domain centred on the optimised parameters. Alternatively uncertainties can be estimated by the Ceres solver which uses the gradient of the chi squared error at the optimised point. 

Quality of the uncertainty estimates can be roughly predicted by the reduced chi squared,  $\chi^2_\nu = \frac{\chi^2}{d.o.f} \approx 1$.  Orders of magnitude smaller than this and the feasible set would be wide and the gradient would be near zero  implying overestimated experimental uncertainties resulting in a large parameter uncertainty. Larger $\chi_\nu^2$ mean the feasible set is empty and imply experimental uncertainties are underestimated, or fitting is limited by MuDirac's precision. A complete description of the parameter uncertainty accounting for uncertainty in the simulation is more complicated and providing more robust uncertainty estimates is subject for further work. 

The experimental muonic x-ray transition energies used for these \texttt{MuDirac} simulations, as well as the method used for estimating the upper bound for the errors, are in the Supplementary Information. 

\begin{table}[]
\centering
\begin{tabular}{|c|c|c|}
\hline
\textbf{Element}	& \texttt{MuDirac} \textbf{$R_{\textrm{rms}}$ [fm]} &	\textbf{Reference $R_{\textrm{rms}}$\cite{Engfer1974} [fm]} \\ 
\hline
 $\mathrm{^{89}Y}$ &	4.2444$\pm$0.043 & 4.234$\pm$0.007 \\
$\mathrm{^{114}Cd}$ & 4.6096$\pm$0.009 & 4.624 $\pm$0.008 \\
$\mathrm{^{197}Au}$ & 5.4207$\pm$0.02 & 5.415$\pm$0.007 \\
$\mathrm{^{206}Pb}$ & 5.4853$\pm 0.0025$ & 5.4839$\pm$0.028 \\
\hline
\end{tabular}
\caption{\texttt{MuDirac} $R_{\textrm{rms}}$ results for four elements compared to recent reference $R_{\textrm{rms}}$ values from Engfer\cite{Engfer1974} \label{tableRrms}.  The uncertainties in the \texttt{MuDirac} values are upper bounds for the errors. How these upper bounds were estimated is shown in the Supplementary Information. }
\end{table}

\clearpage
\section{Conclusions}
\vspace{5mm}

In this work, we have presented and tested \texttt{MuDirac}(1.3.0), a sustainable software tool that can be used to efficiently calculate the 2pF model parameters c and t using, at least, two experimentally measured muonic X-ray transition energies.  \texttt{MuDirac}(1.3.0) is a publicly available and computationally efficient software tool that is at the disposal of the negative muon community. Provided that accurate experimental data is available, \texttt{MuDirac}(1.3.0)
can be used to accurately estimate nuclear properties such as the nuclear charge
distribution, $\rho$, and the root-mean-square radius $R_{\textrm{rms}}$ of a significant variety of atoms and their isotopes. \texttt{MuDirac}(1.3.0) is available to download and install as an executable.  
\newline

We are currently also working on improving the modelling of the cascading process in \texttt{MuDirac}.  

\section{Acknowledgements }
Thanks to Jess Huntley and the Computational Mathematics Theme in the Scientific Computing Department of the STFC for their discussions on optimization and support with using FitBenchmarking.
\newline

This work was supported by the Ada Lovelace Centre, STFC, UKRI.  Funds were also provided by the European Union programme Horizon Europe (HORIZON-INFRA-2021-EOSC-01-04) under grant agreement number 101057388 and by UK Research and Innovation (UKRI) under the UK government’s Horizon Europe funding guarantee grant number 10038963.

\section{ List author names and the contributions made to the article}

This list uses terms from the NISO Contributor Roles Taxonomy (CRediT) https://credit.niso.org 

\begin{itemize}
    \item \textbf{L. Liborio:}. Writing of the original draft. Conceptualization and formulation of overarching research goals and aims. Supervision and leadership responsibility for the research activity planning and execution. Acquisition of the financial support for the project. Development of the theoretical framework for the manuscript. Validation  of the research outputs. Applied computational tools to analyse data. Contributed to the development of the graphs in the manuscript. 
    \item \textbf{M. Kumar:} Review and edit of the manuscript. Development of software to implement the theoretical framework: the reverse-engineering of the Dirac equation. Optimisation and testing of the code components. Preparation of graphs for data presentation. Application of computational techniques to analyse data. Contributed to discussions of the theoretical framework. Validation  of the research outputs. Data curation: put old experimental research data in modern digital format, which allows for easier verification and re-use.
    \item \textbf{S. Devadasan:} Review and edit of the manuscript. Contributed to the development of software to implement the theoretical framework.  Optimisation and testing of the code components.  Validation  of the research outputs. Supervision  of the software development planning and execution. 
    \item \textbf{P. D. Jones:} Review and edit of the manuscript. Contributed to the discussions on software development. 
    \item \textbf{M. Plummer:} Review and edit of the manuscript. Contributed to the discussions on the theoretical framework. Acquisition of the financial support for the project.
    \item \textbf{A. Hillier:} Review and edit of the manuscript. Acquisition of the financial support for the project. 
    \item \textbf{A. Bartok:} Review and edit of the manuscript. Acquisition of the financial support for the project. 
    
\end{itemize}

\section{Supplementary Information}

\subsection{Muonic atoms datasets for optimising $\chi^2(c,t)$}

The region of the parameter space where bands of feasible Fermi parameters for each transition overlap is defined as $\Omega^*$. In this region, the nuclear model used will fit the experimental muonic X-ray transition energies used to determine it. This reduces the wide range of valid 2pF parameters to a subset which can be described by a optimal value with small uncertainty determined by a least squares method. The optimal parameters $c^*$ and $t^*$ are found by minimising the difference between the experimental and simulated energies, as defined in equations (\ref{eq: SE}) to (\ref{eq: opt 2pF parameters}):

\begin{align}
SE_{k}(c, t) & = {\frac{(E_{k}^{exp}-E_{k}^{sim}(c,t))^2}
{\sigma_{k}^2}}, \label{eq: SE}\\[10pt] 
\chi^2(c, t) & = \displaystyle\sum_{k} {SE}_{k}(c, t) \label{eq: chi sq}, \\[10pt] 
\Omega^* & = \{ c, t \in \Omega : \forall k \in I,\; SE_{k}(c,t) < 1 \}, \label{eq: 2pF valid domain}\\[10pt] 
(c^*, t^*) & = arg\min_{c,t \in \Omega}\chi^2(c,t) \label{eq: opt 2pF parameters},\\[10pt] \notag{}
\end{align}

For a given muonic atom, the variable that is optimised when looking for optimum values $c^{*}$ and $t^{*}$ is $\chi^2(c,t)$, as defined by equation (\ref{eq: chi sq}).  For determining the best optimisation algorithms, we optimised $\chi^2(c,t)$ using muonic atoms datasets taken from Engfer et. al\cite{Engfer1974}. Table (\ref{34atomtable}) shows these muonic atoms datasets. These muonic datasets are counted considering type of atom, isotope and experimental data. For example, $\mathrm{Bi}$ contributes with 3 datasets, as the table contains data obtained from 3 different papers for the $\mathrm{^{209}Bi}$ isotope. $\mathrm{Tl}$ contributes with 2 datasets, as the table contains data obtained for the $\mathrm{^{203}Tl}$ and $\mathrm{^{205}Tl}$ isotopes. 

\begin{longtable}{llllrr}
\caption{X-ray transition energies (keV) grouped by element.}
\label{tab:xray_energies_elements} \\
\hline
Ref. in \cite{Engfer1974} & Element & Isotope & Transition & Energy (keV) & Error (keV) \\
\hline
\endfirsthead

\hline
Ref & Element & Isotope & Transition & Energy (keV) & Error (keV) \\
\hline
\endhead

\hline
\endfoot

\hline
\endlastfoot

\multicolumn{6}{l}{\textbf{Zinc (Zn)}} \\
70JP1 & Zn & 66 & K1--L2 & 1592.970 & 0.400 \\
70JP1 & Zn & 66 & K1--L3 & 1600.150 & 0.400 \\
70JP1 & Zn & 66 & L2--M4 & 360.750 & 0.250 \\
70JP1 & Zn & 66 & L3--M5 & 354.290 & 0.250 \\
70JP1 & Zn & 68 & K1--L2 & 1591.220 & 0.390 \\
70JP1 & Zn & 68 & K1--L3 & 1598.220 & 0.390 \\
70JP1 & Zn & 68 & L2--M4 & 360.590 & 0.250 \\
70JP1 & Zn & 68 & L3--M5 & 354.150 & 0.250 \\

\multicolumn{6}{l}{\textbf{Yttrium (Y)}} \\
70KM1 & Y & 89 & K1--L2 & 2420.120 & 0.440 \\
70KM1 & Y & 89 & K1--L3 & 2439.380 & 0.500 \\
70KM1 & Y & 89 & L2--M4 & 616.380 & 0.400 \\
70KM1 & Y & 89 & L3--M5 & 599.390 & 0.400 \\

\multicolumn{6}{l}{\textbf{Niobium (Nb)}} \\
72PO1 & Nb & 93 & K1--L2 & 2603.420 & 0.090 \\
72PO1 & Nb & 93 & K1--L3 & 2626.600 & 0.080 \\
72PO1 & Nb & 93 & L2--M4 & 683.030 & 0.110 \\
72PO1 & Nb & 93 & L3--M5 & 662.540 & 0.110 \\
71CC1 & Nb & 93 & K1--L2 & 2603.200 & 0.500 \\
71CC1 & Nb & 93 & K1--L3 & 2626.400 & 0.400 \\
71CC1 & Nb & 93 & L2--M4 & 687.170 & 0.080 \\
71CC1 & Nb & 93 & L3--M5 & 665.880 & 0.080 \\

\multicolumn{6}{l}{\textbf{Cadmium (Cd)}} \\
71KB1 & Cd & 114 & K1--L2 & 3223.740 & 0.460 \\
71KB1 & Cd & 114 & K1--L3 & 3263.630 & 0.370 \\
71KB1 & Cd & 114 & L2--M4 & 940.580 & 0.230 \\
71KB1 & Cd & 114 & L3--M5 & 905.850 & 0.260 \\

\multicolumn{6}{l}{\textbf{Indium (In)}} \\
71KB1 & In & 115 & K1--L2 & 3322.670 & 0.420 \\
71KB1 & In & 115 & K1--L3 & 3366.270 & 0.420 \\
71KB1 & In & 115 & L2--M4 & 981.640 & 0.480 \\
71KB1 & In & 115 & L3--M5 & 943.390 & 0.480 \\

\multicolumn{6}{l}{\textbf{Tin (Sn)}} \\
71KB1 & Sn & 116 & K1--L2 & 3419.060 & 0.400 \\
71KB1 & Sn & 116 & K1--L3 & 3464.780 & 0.470 \\
71KB1 & Sn & 116 & L2--M4 & 1022.090 & 0.230 \\
71KB1 & Sn & 116 & L3--M5 & 982.230 & 0.220 \\
71KB1 & Sn & 124 & K1--L2 & 3398.910 & 0.360 \\
71KB1 & Sn & 124 & K1--L3 & 3444.290 & 0.350 \\
71KB1 & Sn & 124 & L2--M4 & 1021.420 & 0.210 \\
71KB1 & Sn & 124 & L3--M5 & 981.690 & 0.210 \\

\multicolumn{6}{l}{\textbf{Tellurium (Te)}} \\
71KB1 & Te & 126 & K1--L2 & 3577.630 & 0.580 \\
71KB1 & Te & 126 & K1--L3 & 3629.260 & 0.420 \\
71KB1 & Te & 126 & L2--M4 & 1105.660 & 0.230 \\
71KB1 & Te & 126 & L3--M5 & 1060.660 & 0.220 \\

\multicolumn{6}{l}{\textbf{Iodine (I)}} \\
71KB1 & I & 127 & K1--L2 & 3667.350 & 0.200 \\
71KB1 & I & 127 & K1--L3 & 3723.230 & 0.200 \\
71KB1 & I & 127 & L2--M4 & 1150.420 & 0.150 \\
71KB1 & I & 127 & L3--M5 & 1101.840 & 0.200 \\
66AB1 & I & 127 & K1--L2 & 3667.600 & 3.000 \\
66AB1 & I & 127 & K1--L3 & 3721.600 & 2.500 \\
66AB1 & I & 127 & L2--M4 & 1146.700 & 3.500 \\
66AB1 & I & 127 & L3--M5 & 1098.000 & 3.000 \\

\multicolumn{6}{l}{\textbf{Cesium (Cs)}} \\
69LB1 & Cs & 133 & K1--L2 & 3840.080 & 0.480 \\
69LB1 & Cs & 133 & K1--L3 & 3902.180 & 0.480 \\
69LB1 & Cs & 133 & L2--M4 & 1238.350 & 0.400 \\
69LB1 & Cs & 133 & L3--M5 & 1185.470 & 0.400 \\
66AB1 & Cs & 133 & K1--L2 & 3836.100 & 3.000 \\
66AB1 & Cs & 133 & K1--L3 & 3899.100 & 3.500 \\
66AB1 & Cs & 133 & L2--M4 & 1241.600 & 3.000 \\
66AB1 & Cs & 133 & L3--M5 & 1188.600 & 3.000 \\

\multicolumn{6}{l}{\textbf{Lanthanum (La)}} \\
66AB1 & La & 139 & K1--L2 & 4001.300 & 4.000 \\
66AB1 & La & 139 & K1--L3 & 4071.200 & 4.000 \\
66AB1 & La & 139 & L2--M4 & 1328.400 & 3.000 \\
66AB1 & La & 139 & L3--M5 & 1266.800 & 3.000 \\

\multicolumn{6}{l}{\textbf{Cerium (Ce)}} \\
70JP11 & Ce & 140 & K1--L2 & 4097.610 & 0.130 \\
70JP11 & Ce & 140 & K1--L3 & 4171.240 & 0.110 \\
70JP11 & Ce & 140 & L2--M4 & 1376.410 & 0.090 \\
70JP11 & Ce & 140 & L3--M5 & 1313.630 & 0.080 \\

\multicolumn{6}{l}{\textbf{Praseodymium (Pr)}} \\
66AB1 & Pr & 141 & K1--L2 & 4184.300 & 5.000 \\
66AB1 & Pr & 141 & K1--L3 & 4258.800 & 5.500 \\
66AB1 & Pr & 141 & L2--M4 & 1422.600 & 3.000 \\
66AB1 & Pr & 141 & L3--M5 & 1356.700 & 3.000 \\
69LB1 & Pr & 141 & K1--L2 & 4184.800 & 0.340 \\
69LB1 & Pr & 141 & K1--L3 & 4262.370 & 0.340 \\
69LB1 & Pr & 141 & L2--M4 & 1424.220 & 0.450 \\
69LB1 & Pr & 141 & L3--M5 & 1358.610 & 0.450 \\

\multicolumn{6}{l}{\textbf{Neodymium (Nd)}} \\
70MB1 & Nd & 142 & K1--L2 & 4270.440 & 0.500 \\
70MB1 & Nd & 142 & K1--L3 & 4352.060 & 0.400 \\
70MB1 & Nd & 142 & L2--M4 & 1472.320 & 0.250 \\
70MB1 & Nd & 142 & L3--M5 & 1402.970 & 0.250 \\
70MB1 & Nd & 144 & K1--L2 & 4254.880 & 0.550 \\
70MB1 & Nd & 144 & K1--L3 & 4336.380 & 0.450 \\
70MB1 & Nd & 144 & L2--M4 & 1471.820 & 0.250 \\
70MB1 & Nd & 144 & L3--M5 & 1402.590 & 0.250 \\
70MB1 & Nd & 146 & K1--L2 & 4240.500 & 0.500 \\
70MB1 & Nd & 146 & K1--L3 & 4321.180 & 0.400 \\
70MB1 & Nd & 146 & L2--M4 & 1470.990 & 0.250 \\
70MB1 & Nd & 146 & L3--M5 & 1402.580 & 0.200 \\
70MB1 & Nd & 148 & K1--L2 & 4223.760 & 0.500 \\
70MB1 & Nd & 148 & K1--L3 & 4303.460 & 0.450 \\
70MB1 & Nd & 148 & L2--M4 & 1470.970 & 0.250 \\
70MB1 & Nd & 148 & L3--M5 & 1403.540 & 0.250 \\
70MB1 & Nd & 150 & K1--L2 & 4198.300 & 0.400 \\
70MB1 & Nd & 150 & K1--L3 & 4266.370 & 0.500 \\
70MB1 & Nd & 150 & L2--M4 & 1474.880 & 0.250 \\
70MB1 & Nd & 150 & L3--M5 & 1419.080 & 0.250 \\

\multicolumn{6}{l}{\textbf{Gold (Au)}} \\
66AB1 & Au & 197 & K1--L2 & 5592.800 & 5.000 \\
66AB1 & Au & 197 & K1--L3 & 5762.500 & 5.000 \\
66AB1 & Au & 197 & L2--M4 & 2474.400 & 2.000 \\
66AB1 & Au & 197 & L3--M5 & 2343.100 & 2.500 \\

\multicolumn{6}{l}{\textbf{Thallium (Tl)}} \\
72BE1 & Tl & 203 & K1--L2 & 5726.140 & 0.670 \\
72BE1 & Tl & 203 & K1--L3 & 5906.380 & 0.670 \\
72BE1 & Tl & 203 & L2--M4 & 2587.730 & 0.190 \\
72BE1 & Tl & 203 & L3--M5 & 2448.220 & 0.190 \\
72BE1 & Tl & 205 & K1--L2 & 5717.210 & 0.650 \\
72BE1 & Tl & 205 & K1--L3 & 5897.290 & 0.670 \\
72BE1 & Tl & 205 & L2--M4 & 2586.290 & 0.200 \\
72BE1 & Tl & 205 & L3--M5 & 2446.560 & 0.190 \\

\multicolumn{6}{l}{\textbf{Lead (Pb)}} \\
69AH2 & Pb & 206 & K1--L2 & 5788.330 & 0.480 \\
69AH2 & Pb & 206 & K1--L3 & 5973.980 & 0.440 \\
69AH2 & Pb & 206 & L2--M4 & 2643.750 & 0.360 \\
69AH2 & Pb & 206 & L3--M5 & 2501.450 & 0.430 \\
66AB1 & Pb & 206 & K1--L2 & 5786.900 & 5.000 \\
66AB1 & Pb & 206 & K1--L3 & 5972.300 & 5.000 \\
66AB1 & Pb & 206 & L2--M4 & 2643.200 & 3.000 \\
66AB1 & Pb & 206 & L3--M5 & 2500.600 & 1.500 \\
68PO1 & Pb & 206 & K1--L2 & 5789.000 & 2.800 \\
68PO1 & Pb & 206 & K1--L3 & 5974.100 & 2.000 \\
68PO1 & Pb & 206 & L2--M4 & 2645.500 & 2.300 \\
68PO1 & Pb & 206 & L3--M5 & 2503.400 & 2.300 \\
69AH2 & Pb & 208 & K1--L2 & 5778.930 & 0.500 \\
69AH2 & Pb & 208 & K1--L3 & 5963.770 & 0.450 \\
69AH2 & Pb & 208 & L2--M4 & 2641.480 & 0.420 \\
69AH2 & Pb & 208 & L3--M5 & 2500.070 & 0.450 \\
71JP1 & Pb & 208 & K1--L2 & 5778.500 & 0.500 \\
71JP1 & Pb & 208 & K1--L3 & 5963.300 & 0.500 \\
71JP1 & Pb & 208 & L2--M4 & 2641.940 & 0.200 \\
71JP1 & Pb & 208 & L3--M5 & 2500.340 & 0.190 \\

\multicolumn{6}{l}{\textbf{Bismuth (Bi)}} \\
67BC2 & Bi & 209 & K1--L2 & 5841.500 & 3.000 \\
67BC2 & Bi & 209 & K1--L3 & 6032.400 & 3.000 \\
67BC2 & Bi & 209 & L2--M4 & 2699.500 & 1.000 \\
67BC2 & Bi & 209 & L3--M5 & 2552.800 & 1.000 \\
66AB1 & Bi & 209 & K1--L2 & 5839.700 & 5.500 \\
66AB1 & Bi & 209 & K1--L3 & 6032.200 & 5.000 \\
66AB1 & Bi & 209 & L2--M4 & 2700.200 & 2.500 \\
66AB1 & Bi & 209 & L3--M5 & 2554.800 & 2.000 \\
68PO1 & Bi & 209 & K1--L2 & 5843.200 & 2.500 \\
68PO1 & Bi & 209 & K1--L3 & 6034.000 & 2.200 \\
68PO1 & Bi & 209 & L2--M4 & 2702.800 & 2.500 \\
68PO1 & Bi & 209 & L3--M5 & 2556.100 & 2.500 \\
\label{34atomtable}
\end{longtable}

\section{Experimental muonic x-ray transition energies used for the \texttt{MuDirac} simulations in table (\ref{tableRrms}). }
The experimental muonic x-ray transition energies were taken from Engfer et. al.\cite{Engfer1974}. The energies are in eV.
\newline

$\mathrm{^{89}Y}$ experimental input file <name>.xr.in 
\begin{Verbatim}[commandchars=\\\{\}]
\end{Verbatim}
\begin{center}
\begin{tabular}{llllll}
xr$\_$lines: & $K1-L2$ & $K1-L3$ & $L2-M4$ & $L3-M5$\\
xr$\_$energy: & 2420120 & 2439380 & 616380 & 599390 \\
xr$\_$error: & 440 & 500 & 400 & 400\\
\end{tabular}
\end{center}

$\mathrm{^{114}Cd}$ experimental input file <name>.xr.in 
\begin{Verbatim}[commandchars=\\\{\}]
\end{Verbatim}
\begin{center}
\begin{tabular}{llllll}
xr$\_$lines: & $K1-L2$ & $K1-L3$ & $L2-M4$ & $L3-M5$\\
xr$\_$energy: & 3223740 & 3263630 & 940580 & 905850 \\
xr$\_$error: & 460 & 370 & 230 & 260\\
\end{tabular}
\end{center}

$\mathrm{^{197}Au}$ experimental input file <name>.xr.in 
\begin{Verbatim}[commandchars=\\\{\}]
\end{Verbatim}
\begin{center}
\begin{tabular}{lllll}
xr$\_$lines: & $K1-L2$ & $K1-L3$ & $L2-M4$ & $L3-M5$ \\
xr$\_$energy: & 5592800 & 5762500 & 2474400 & 2343100 \\
xr$\_$error: & 5000 & 5000 & 2000 & 2500 \\
\end{tabular}
\end{center}

$\mathrm{^{206}Pb}$ experimental input file <name>.xr.in 
\begin{Verbatim}[commandchars=\\\{\}]
\end{Verbatim}
\begin{center}
\begin{tabular}{lllll}
xr$\_$lines: & $K1-L2$ & $K1-L3$ & $L2-M4$ & $L3-M5$ \\
xr$\_$energy: & 5788330 & 5973980 & 2643750 & 2501450 \\
xr$\_$error: &  480 & 440 & 360 & 430 \\
\end{tabular}
\end{center}

\subsection{ Estimation of the upper bound for the uncertainties in the nuclear radii shown in table (\ref{tableRrms}). }

\begin{table}[hb]
\centering
\begin{tabular}{|c|c|c|}
\hline
\textbf{Element}	& \texttt{MuDirac} \textbf{$R_{rms}$ [fm]} &	\textbf{Reference $R_{rms}$\cite{Engfer1974} [fm]} \\ 
\hline
 $\mathrm{^{89}Y}$ &	4.2444$\pm$0.043 & 4.234$\pm$0.007 \\
$\mathrm{^{114}Cd}$ & 4.6096$\pm$0.009 & 4.624 $\pm$0.008 \\
$\mathrm{^{197}Au}$ & 5.4207$\pm$0.02 & 5.415$\pm$0.007 \\
$\mathrm{^{206}Pb}$ & 5.4853$\pm 0.0025$ & 5.4839$\pm$0.028 \\
\hline
\end{tabular}
\caption{\texttt{MuDirac} $R_{rms}$ results for four elements compared to recent reference $R_{rms}$ values from Engfer\cite{Engfer1974} \label{tableRrms}.  The uncertainties in the \texttt{MuDirac} values are upper bounds for the errors. }
\end{table}

\begin{figure}[!htb]
    \centering
        \includegraphics[width=0.9\textwidth]{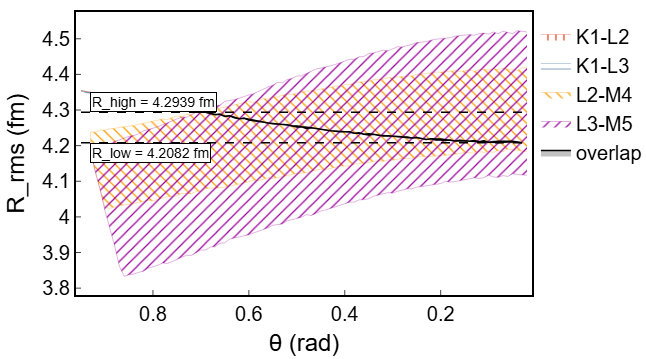}
    \caption{2pF nuclear model parameter bands, which reproduce valid muonic X-ray energies for four selected transitions in $^{89}\text{Y}$, shown in polar coordinates.  The upper and lower bounds for the allowed values of $R_{rms}^2$ are indicated with dashed lines. \label{fig:89Y}}
\end{figure}

\begin{figure}[!htb]
    \centering
        \includegraphics[width=0.9\textwidth]{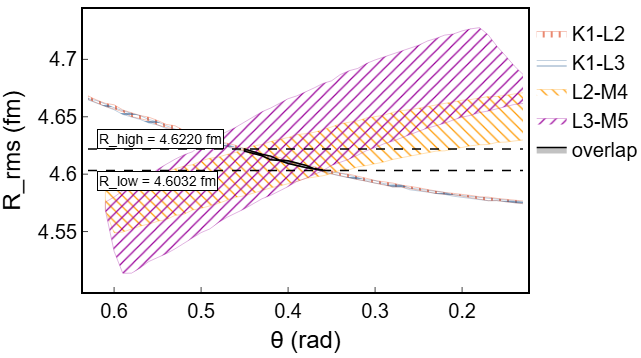}
    \caption{2pF nuclear model parameter bands, which reproduce valid muonic X-ray energies for four selected transitions in $^{114}\text{Cd}$, shown in polar coordinates.  The upper and lower bounds for the allowed values of $R_{rms}^2$ are indicated with dashed lines. \label{fig:114Cd}}
\end{figure}

\begin{figure}[!htb]
    \centering
        \includegraphics[width=0.9\textwidth]{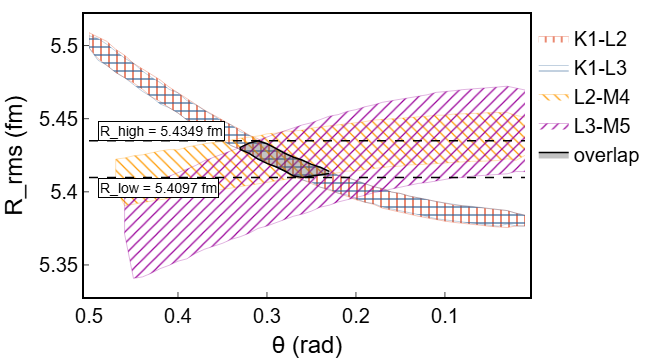}
    \caption{2pF nuclear model parameter bands, which reproduce valid muonic X-ray energies for four selected transitions in $^{197}\text{Au}$, shown in polar coordinates.  The upper and lower bounds for the allowed values of $R_{rms}^2$ are indicated with dashed lines. \label{fig:197Au}}
\end{figure}

\begin{figure}[!htb]
    \centering
        \includegraphics[width=0.9\textwidth]{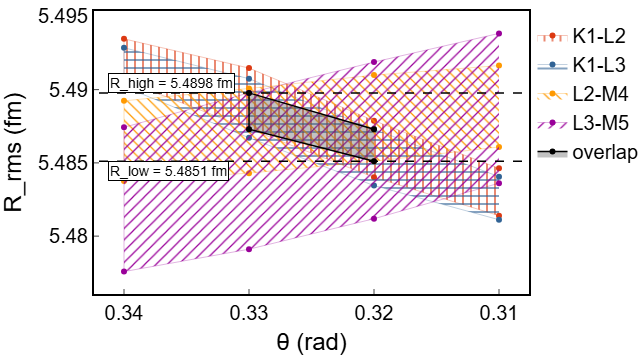}
    \caption{2pF nuclear model parameter bands, which reproduce valid muonic X-ray energies for four selected transitions in $^{206}\text{Pb}$, shown in polar coordinates.  The upper and lower bounds for the allowed values of $R_{rms}^2$ are indicated with dashed lines. \label{fig:206Pb}}
\end{figure}

Figures (\ref{fig:89Y}), (\ref{fig:114Cd}), (\ref{fig:197Au}) and (\ref{fig:206Pb}) show the 2pF parameters' domains for $^{89}\text{Y}$, $^{114}\text{Cd}$. $^{197}\text{Au}$ and $^{206}\text{Pb}$ in polar coordinates.  The polar coordinate system better illustrates the model dependence of the $R_{rms}$ associated with each band. Hence, the model solutions for each band do not perfectly coincide, which results in a more restricted overlap region for the 2pF parameters that reproduce the experimentally observed transition energies. The figures also show the upper and lower bounds for the allowed values of $R_{rms}^2$ for $^{89}\text{Y}$, $^{114}\text{Cd}$, $^{197}\text{Au}$ and $^{206}\text{Pb}$.  Each $R_{rms}^2$ value estimated by \texttt{MuDirac} cannot be outside of these ranges. 

\FloatBarrier
\printbibliography

\end{document}